\documentclass[onecolumn, tighten]{aastex631}
\usepackage{nicefrac}
\usepackage{microtype}
\usepackage{booktabs}
\usepackage{rotating}
\usepackage{xcolor}
\usepackage[utf8]{inputenc}
\usepackage{amsmath}

\DeclareUnicodeCharacter{2212}{\ensuremath{-}}

\received{?}
\revised{?}
\accepted{?}
\submitjournal{Frontiers}
\shorttitle{MHD Simulations of Exoplanet Magnetosphere}
\shortauthors{Bagheri et al.}

\begin{document}

\title{A Fresh Look into the Interaction of Exoplanets Magnetosphere with Stellar Winds using MHD Simulations}

\author{Fatemeh Bagheri}
\affil{NASA Goddard Space Flight Center, Greenbelt, MD, USA}
\affil{Department of Physics, University of Texas at Arlington, Arlington, TX, USA}
\correspondingauthor{Fatemeh Bagheri}
\email{fatemeh.bagheri@nasa.gov}

\author{Ramon E. Lopez}
\affil{Department of Physics, University of Texas at Arlington, Arlington, TX, USA}

\author{Kevin Pham}
\affil{High Altitude Observatory (HAO), Boulder, CO, USA}

\begin{abstract}

%%% Leave the Abstract empty if your article does not require one, please see the Summary Table for full details.

\noindent Numerous numerical studies have been carried out in recent years that simulate different aspects of exoplanets' magnetosphere and stellar winds. These studies have focused primarily on hot Jupiters with sun-like stars. This study addresses the challenges inherent in utilizing existing MHD codes to model hot Jupiter-star systems. Due to the scaling of the system and the assumption of a uniformly flowing stellar wind at the outer boundary of the simulation, MHD codes necessitate a minimum distance of greater than 0.4 au for a Jupiter-like planet orbiting a sun-like star to avoid substantial violations of the code's assumptions. Additionally, employing the GAMERA (Grid Agnostic MHD for Extended Research Applications) MHD code, we simulate star-planet interactions considering various stellar types (Sun-like and M Dwarf stars) with both Jupiter-like and Earth-like planets positioned at varying orbital distances. Furthermore, we explore the impact of tidal locking on the total power within the magnetosphere-ionosphere systems.

%\tiny
 %\keyFont{ \section{Keywords:} Exoplanet, MHD Simulations, AKR Emission, Planetary Magnetic field, Close-in Exoplanet} %All article types: you may provide up to 8 keywords; at least 5 are mandatory.
\end{abstract}

\section{Introduction} \label{sec:intro}

\noindent The study of exoplanets has emerged as a captivating frontier in astrophysics, providing unprecedented insights into the diversity and potential habitability of distant worlds. The discovery of the first exoplanets \citep{mayor1995jupiter} has revolutionized our understanding of planetary systems and expanded the scope of the search for extraterrestrial life. Advancements in the exoplanet detection techniques, such as the transit method, radial velocity measurements, microlensing, and direct imaging have led to the discovery of thousands of exoplanets and characterized their properties with increasing precision \citep{wright2012frequency, knutson2014featureless, mandel2002analytic, seager2003unique, gaudi2012microlensing, bagheri2019detection, traub2010direct, guyon2005exoplanet}. So far, more than 5000 exoplanets have been detected, presenting a rich tapestry of planetary populations that help find habitable planets. Habitability, defined as the conditions suitable for life as we know it on Earth, encompasses a complex interplay of factors ranging from a planet's size, composition, and atmosphere to its orbital characteristics and the nature of its host star. While Earth remains the sole known haven for life in the cosmos, the prospect of discovering habitable environments beyond our solar system has been an ongoing research topic.\\
\\
In the quest to unravel the mysteries of exoplanetary habitability, one fundamental and often overlooked factor emerges as a critical player: the magnetic field. On Earth, the geomagnetic field acts as a protective shield, deflecting charged particles from the solar wind and cosmic rays that could otherwise strip away our atmosphere and erode the conditions conducive to life \citep{griebetameier2005cosmic}. Although it was previously believed that Magnetic fields shield against stellar winds and influence atmospheric gas escape rates, recent studies show that a magnetosphere clearly impacts ionospheric outflow. An intrinsic magnetic field is not necessarily needed to prevent atmospheric erosion \citep{ramstad2021intrinsic, gronoff2020atmospheric}. Rather, it is a combination of the planet's magnetic field and stellar wind, as well as the characteristic of the planetary atmosphere, that represents a delicate equilibrium that governs habitability. Magnetic fields not only shield against stellar winds but also influence the escape rates of atmospheric gases. It can regulate the long-term stability of a planet's atmosphere if field strength is strong enough relative to the upstream stellar wind and EUV flux from the host star and the gravity of the planet \citep{lundin2007planetary, ramstad2021intrinsic}. This, in turn, influences surface temperatures and the potential for liquid water to exist, shaping the very foundation of habitability. \\
%While the pursuit of habitable environments beyond our solar system has traditionally focused on parameters such as a planet's proximity to its host star and the composition of its atmosphere, understanding the presence, strength, and dynamics of magnetic fields becomes paramount in the quest to identify potentially habitable worlds. 
\noindent One possible way to study exoplanets' magnetic fields is by observing their radio emission.  Within our solar system, Jupiter stands out as a prominent radio emitter, producing powerful bursts of radiation originating from its intense magnetic field. However, all magnetized planets in our solar system, even seemingly less dynamic planets, such as Earth and Saturn, contribute their own distinct radio signatures, each revealing the unique intricacies of their magnetospheric environments. Like Earth, Jupiter and Saturn have auroras, which are bright and dynamic displays of light near the polar regions caused by the interaction of charged particles with the planet's atmosphere. Within the auroral zones, the magnetic fields trap energetic electrons. Some of these electrons can undergo a process called the Electron Cyclotron Maser Instability (ECMI). This process amplifies electromagnetic waves in the kilometric range. The amplified waves manifest as radio emission known as Auroral Kilometric Radiation (AKR). The emitted radio waves are detected at kilometric wavelengths, typically ranging from a few hundred kilohertz to a few megahertz. The exploration of AKR presents a promising avenue for directly detecting exoplanets, offering an independent means of inquiry distinct from their influence on host stars. The extension of AKR detection methods holds the potential for direct exoplanet detection and offers a unique opportunity to unravel the origins of exoplanetary magnetic fields.\\

\noindent Ground-based observatories and spaceborne instruments, such as the Voyager and Cassini spacecraft, have provided crucial data, enabling scientists to probe the radio signals emanating from the gas giants and icy moons of our solar system \citep{asmar2021solar}. Beyond these familiar realms, the detection of radio emissions from exoplanets promises to unveil new dimensions of planetary diversity and magnetospheric complexity. In recent decades, advances in radio telescopes, such as the Square Kilometre Array (SKA) telescope, have extended our capacity to explore and even detect planetary radio emissions beyond our immediate neighborhood. \\ 

\noindent Two Priority Science Question Topics in the Origins, Worlds, and Life Planetary Science \& Astrobiology Decadal Survey report have elements involving the existence of planetary magnetic fields and their interactions with the solar wind, Q6: "Solid body atmospheres, exospheres, magnetospheres, and climate evolution" and Q12.7: "Exoplanets, Giant planet structure and evolution" \citep{national2022origins}. To answer these questions, in this paper, we study the interaction of the exoplanets' magnetosphere and stellar winds using Magneto-Hydro-Dynamic(MHD) simulations. Field-aligned currents (FACs) are identified as the primary and most effective sources of planetary ECMI emission \citep{treumann2006electron}. Therefore, this paper aims to estimate FACs and the total power within the magnetosphere of two types of exoplanets under different conditions. We use a variant of MHD numerical simulations known as GAMERA \citep{zhang2019gamera}. We determine the overall power within the planetary magnetosphere by accounting for various stellar types, the influence of tidal locking and exoplanet rotation, and different orbital distances. Our approach involves conducting GAMERA simulations to explore different scenarios of the star-planet interaction, including a Sun-like star with both a rocky and a Jovian planet and an M Dwarf star with similar planetary configurations. The GAMERA simulation outputs ionospheric power data for each scenario based on stellar wind parameters, orbital distance, and conductance within the exoplanet's ionosphere. The paper's structure is outlined as follows: Section 2 details the methodology for the MHD simulations, while Section 3 summarizes the results. Section 4 discusses the implications of these findings and presents the conclusion.

\section{Magneto-hydro-dynamic Simulations and their Limitations in Exoplanet Sciences}

\noindent Based on our knowledge of the solar system planets, the average ECMI is related to the total solar wind power and the magnetic field of the planet by using the RBL model \citep{zarka2001magnetically, zarka2004fast, zarka2007plasma}. However, the auroral radio emission strongly correlates with solar wind fluctuations (density, velocity, and/or ram pressure) \citep{gallagher1981correlations, desch1982evidence, genova1989jupiter, bagheri2022solar}. Regardless of its origins, the existence of this correlation indicates that different types of stellar winds can significantly change the power of the auroral radio emission \citep{griessmeier2007exoplanetary}. For example, planets around young stars may be more luminous in radio emission than the solar system planets \citep{wood2005new}. More realistic estimates of the planetary radio emission are possible via MHD simulations of the interactions between the stellar wind and the planetary magnetic field. Global MHD simulations have been extensively used to study magnetospheric dynamics and ionosphere phenomena for the Earth and other solar system planets since the late 70s and early 80s. In recent years, a few studies have used MHD simulations to study the magnetosphere of exoplanets, especially close-in exoplanets. Close-in planets (semimajor axis $a \leq 0.05$ au) constitute a special subset of the exoplanetary population. Since it is unclear whether in-situ formation occurs, these planets' current orbital and physical characteristics provide essential constraints on their past evolution and formation process \citep{jackson2008tidal}. The close-in exoplanets with orbital periods $P < 30$ days can be classified into two major categories: 1) hot Jupiters, gas giant planets and mass ranges of $0.2 M_\otimes < M_p < 8 M_\otimes$, and 2) hot Neptunes or super-Earth planets, with a mass $0.008~M_\otimes < M_p < 0.08 M_\otimes$, where $M_\otimes$ is the mass of Jupiter. The close-in exoplanets were unknown until the detection of 51 Peg b (a hot Jupiter) as they do not exist in the Solar system \citep{mayor1995jupiter}. While ECMI is commonly studied in solar system planets, hot Jupiters' ECMI is still a topic of ongoing research. Theoretically, the conditions necessary for ECMI can be met in hot Jupiters. These conditions include a sufficiently strong magnetic field, a population of high-energy electrons, and a free energy source to drive the instability. Hot Jupiters have been observed to possess magnetic fields, and it is believed that they can generate energetic electrons through various mechanisms, such as magnetic reconnection or particle acceleration by plasma waves. This emission occurs when the electrons gain energy from the wave, producing coherent radio waves. The knowledge of the planetary magnetic moment is essential to estimate exoplanetary radio flux. However, the exoplanetary magnetic fields are still unconstrained by observations. There have been numerous estimates of the radio powers generated by exoplanetary auroral emissions due to electron cyclotron masers (e.g., \citet{laneuville2020magnetic}). Still, the lack of detections to date likely reflects the limited sensitivity of many current telescopes and the relatively high frequencies observed compared to the radio frequencies at which solar system planets emit. \citet{cauley2019magnetic} report the derivation of the magnetic field strengths of four hot Jupiter systems, using the power observed in calcium II K emission modulated by magnetic star–planet interactions. They find that the surface magnetic field values for those four hot Jupiters range from 20 G to 120 G, around 10–100 times larger than the values predicted by dynamo scaling laws for planets with rotation periods of around 2–4 days. Even having information on the planetary magnetic moment will not lead to an accurate estimation of the radio emission from the planet using the RBL model. The simplicity of the RBL model has some advantages. Still, it does not provide a complete picture of all processes involved, and its application to exoplanets is extrapolated over many orders of magnitude. There are other factors, such as saturation of the merging rate between exoplanet magnetic field and the stellar wind IMF, which would limit the transfer of energy from stellar wind to the magnetosphere \citep{lopez2016integrated, bagheri2022solar}, that should be taken into account. \citet{nichols2016stellar} used an analytic model to consider an Earth-type Dungey cycle process of magnetic reconnection. They predict that auroral radio emission from hot Jupiters can be two orders of magnitude smaller than RBL-model predictions due to the saturation of the ionosphere. Furthermore, \citep{koskinen2013escape} and \citep{weber2017expanded} mentioned that radio emission might have a problem escaping from the exoplanet because its ionosphere would block the radiation. This is due to another essential condition for ECMI emission, which necessitates the plasma frequency to be lower than the cyclotron frequency, $f_p  < 0.4 f_c$ \citep{griessmeier2007predicting}. In other words, a relatively depleted and strongly magnetized plasma at the source is needed to produce ECMI \citep{zarka2018jupiter}.\\

\noindent The magnetosphere structure of close-in exoplanets can also be different. All magnetized planets in our solar system have a magnetosphere with a structure similar to Earth's magnetosphere. They are characterized by the following main elements: a bow shock, transition region, magnetopause, radiation belts, and a magnetospheric tail. The similarity of the magnetosphere in all magnetized planets in our solar system is because they all are located in the super-Alfv\'enic region of the solar wind since the Alfv\'en radius in the solar wind is $a = 0.1$ au = $22 ~R_\odot$. Close-in exoplanets, with their proximity to their host stars, experience intense irradiation and extreme thermal conditions so that they exhibit complex and possibly different magnetosphere structures than Earth's. Close-in exoplanets might be located in the sub-Alfv\'en surface, where the magnetic pressure exceeds the dynamical pressure, leading to a magnetically dominated region \citep{zhilkin2019possible}, similar to the situation at Ganymede \citep{jia2021magnetosphere}.\\

\noindent For close-in exoplanets, the tidal force and XUV-driven heating are strong contributors to drive the expansion of the planetary atmosphere beyond the Roche lobe so that there is a gaseous open envelope around the planet, formed in the presence of outflows from the nearest Lagrange points. Therefore, close-in exoplanets can experience significant hydrodynamic escapes \citep{shaikhislamov2020three, johnstone2018upper, johnstone2019extreme}. Hydrodynamic escape has been observed for exoplanets close to their host star, including the hot Jupiters HD 209458b. Results in \citep{weber2017expanded} show that high ionospheric plasma densities in hydrodynamically expanded upper atmospheres of close-in extrasolar giant planets prevent the radio emission from escaping the source region or render the generation of radio waves via the ECMI mechanism impossible. \\

\begin{figure}[ht]
%\plotone{qkpic04au.png}
\centering
\includegraphics[width=120mm]{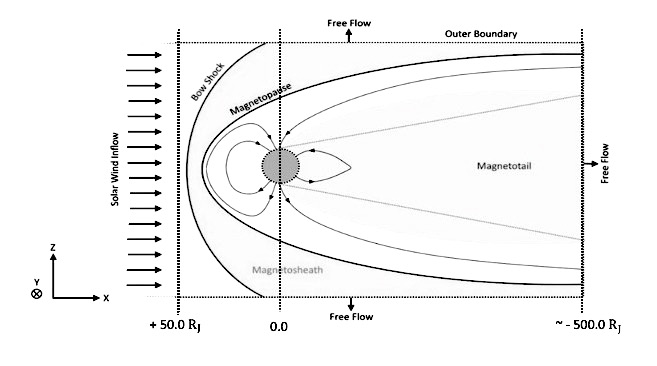}

\caption{A Schematic view of the boundaries in the MHD codes for a Jupiter-like planet with the same size and magnetic field strength as Jupiter.
\label{fig:MHDbox}}
\end{figure}

\noindent Despite efforts to use MHD codes for simulating the magnetospheres of exoplanets orbiting close to their host stars [e.g., \cite{turnpenney2020magnetohydrodynamic}], MHD codes frequently encounter inherent gaps and limitations, often compensated for through empirical fitting or simplifying assumptions. These empirical constraints and other assumptions derived from the properties of Earth or planets within the solar system may not be universally applicable to other planetary systems. As a result, modeling the plasma environments of exoplanets poses notable challenges to conventional MHD models. A crucial feature in any MHD simulation is how boundary conditions are treated. In MHD codes, the outer boundaries include the upstream, side, and downstream. The upstream boundary where the solar wind enters the simulation is placed upstream from the planet (in the case of Earth at 1 au, about 30 Earth radii ($R_\oplus$), or in the case of Jupiter, about 50 Jupiter radii ($R_\otimes$) equivalent to $0.024$ au), where Dirichlet boundary conditions are specified on the solar wind (SW) and interplanetary magnetic field (IMF) variables (Figure \ref{fig:MHDbox}). 
Therefore, simulations of Jupiter-like planets with orbital distances smaller than $\approx 0.03$ au with sun-like hosts are impossible with any current 3D MHD codes because, for the smaller orbital distances, the star would be on a grid. The basic assumptions of the MHD code would be severely violated.\\%Even if the star is not on the grid, the assumption of a uniform stellar wind flowing in the dayside boundary of the grid would be incorrect since the alignment between flow along the planet-star line and the actual radial flow of the stellar wind depends on being sufficiently far from the star.\\

\noindent The downstream boundary is sufficiently far so that plasma flow velocity has become super-Alfv\'enic at the back of the box to avoid reflections off the back boundary. The side boundary conditions are just the solar wind variables propagated to the corresponding star-ward direction by the solar wind velocity. These conditions are usually in the form of Neumann boundary conditions, i.e., $\frac{\partial\epsilon}{\partial n} = 0$ where $n$ is the coordinate normal to the boundary and $\epsilon$ is any of the MHD variables \citep{xi2015poynting}. In the case of a Jupiter-like planet with the same size and magnetic field as Jupiter, the downstream boundary is at about 500 Jupiter radii from the planet, equivalent to $0.24$ au from the planet. The variation of the solar wind density is proportional to $r^{-2}$. Hence, if the planet were located at the orbital distance $0.3$ au, the solar wind density would be changed by more than 70\% from the upstream boundary to the downstream boundary. In general, for a Jupiter-like planet with the same or greater magnetic field, if the orbital distance is smaller than about $0.4$ au, the solar wind density cannot be assumed as a constant parameter on the boundary of the MHD grid. Similar outstanding issues exist with other solar wind state variables, particularly the magnetic field in existing MHD codes. \\

\noindent Beyond the Alfv\'en surface, where the stellar flow is super Alfv\'enic, MHD waves generated at the planet cannot flow upstream. Instead, the solar wind generates a bow shock when encountering planetary obstacles. This bow shock decelerates and redirects the solar wind, influencing its interaction with the magnetosphere-ionosphere system of the planet \citep{garcia2023star}. On the other hand, within the Alfv\'en surface, the transfer of energy can occur in both directions, towards and away from the star, enabling direct interaction between the planet and its host star [e.g., \cite{cohen2014magnetospheric, saur2013magnetic}]. Therefore, including the stellar wind magnetic field formally leads to a change in the regime for the wind flowing around the hot Jupiter from supersonic to subsonic \citep{zhilkin2019possible}. As a result, no bow shock should form in front of the atmosphere in the subsonic regime \citep{ip2004star}; i.e., the wind flowing around the planet should have a shockless character (for example, TRAPPIST-1e \citep{harbach2021stellar} or AU Mic b \citep{cohen2022space}). Analyses show that many typical hot Jupiters should have shockless-induced magnetospheres \citep{zhilkin2019possible}, [e.g., the magnetosphere of Ganymede \citep{jia2021magnetosphere}]. However, for close-in planets that possess high Keplerian velocities and are frequently located at regions where the host star’s wind is still accelerating, a shock may develop in the direction of the planet's orbital motion(for example, WASP-12b \citep{vidotto2011shock}). \\ 

\noindent \citet{cohen2011dynamics} use an MHD stellar corona model to predict the ambient coronal radio emission and its modulations induced by a close planet on its host star. This work focuses on studying the effect of a close-in planet on the corona. Basically, the exoplanet in this simulation is considered as an additional boundary condition in the corona model \citep{cohen2014magnetospheric}. Therefore, the number of grids corresponding to the planet's size in this model needs to be increased to capture all the magnetospheric/ionospheric features of the planet. Hence it cannot be used to calculate the auroral radio emission from the planet. \\

%\citep{zhang2019gamera}.
%\subsection{Planetary Parameters}
%\section{Results}

\section{An Exoplanet with a Sun-like star}

\noindent The closest feasible orbital distance for the planet to adhere to the assumptions within MHD codes is approximately at the orbit of Mercury (about 0.4 au). The key alteration involves the side boundary conditions occurring over a range of $170 R_p$, stretching from the box's front end to the magnetotail's effective end. For a Jovian scenario (where $R_p$ is roughly equivalent to the radius of Jupiter), this equates to a distance of 0.07 au, resulting in a density change of approximately 35\%. While this adjustment is relatively minor, the discrepancy between a constant side boundary density and one that fluctuates by 35\% should not significantly impact the solution. This difference primarily represents a shift in the lobe field strength of around 25\%, assuming thermal pressure equilibrium and a proton temperature radial profile that is notably flatter than adiabatic \cite{freeman1985cold}. The grid structure operates in a 3D Cartesian layout within the computational space, with the quadrature resolution set to 96 $\times$ 96 $\times$ 128 $~R_p$ for grid spacing. In the Jovian case, the inner and sunward outer radius of the ionosphere are designated as 5 and 50 $R_p$, respectively, while the tail's outer radius is 537.52 $R_p$, and the Low-latitude boundary condition is positioned at 24.09 $R_p$. Although the simulation grid extends well beyond this -120 $R_p$ distance in the X-direction, beyond the effective length of the tail, any nonphysical structures or MHD disturbances generated by the errors in the side boundary conditions would be swept out the back end of the simulation grid by the superalfv\`enic flow. Therefore, we can have confidence that the simulation reasonably represents reality. Additionally, at a distance of 0.4 au, the y-component of the solar wind velocity reaches a maximum of $0.17 \times v_x$ at the grid, indicating that the assumption of solar wind flow in the X-direction remains reasonably intact.
\\ 
\\
\noindent For $0.4$ au, we use solar wind values based on actual measurements of the solar wind at the location of Mercury \citep{diego2020properties}. The typical values for solar wind parameters are, IMF $\approx -32~ \text{nT}$, solar wind temperature $\approx 0.2 \times 10^{6}~\text{K}$, solar wind speed $405 ~\text{km/s}$, and solar wind density $44~ \text{{cc}}^{-1}$. We consider a Jovian planet with few differences from the actual case. We assume that Jupiter's dipole moment aligns with its planetary rotation axis and is directed southward, similar to Earth. Other planetary parameters, including mass, radius, magnetic field strength, angular velocity, and the fraction of helium in the ionosphere, are consistent with those of Jupiter (mass: $1.9 \times 10^{27}$ kg, radius: $71, 492$ km, equatorial magnetic field strength: $426.400$ nT, angular velocity: $9$ hours, and helium fraction: $10\%$). For a rocky planet scenario, the planetary characteristics such as mass, radius, magnetic field strength, angular velocity, and helium fraction in the ionosphere are modeled after Earth's parameters (mass: $5.9722 \times 10^{24}$ kg, radius: $6378.14$ km, equatorial magnetic field strength: $0.2961737$ Gauss, angular velocity: $24$ hours, and helium fraction: $4\%$). \\
\\
Throughout all simulations presented in this study, we maintain a constant ionospheric Pedersen conductance. To determine the Pedersen conductance at varying orbital distances, we employ  
\begin{equation}
\Sigma_P = \kappa \big(\frac{d}{1 au}\big)^\lambda ~\big(\frac{B_J}{B_p}\big) ~\big(\frac{L_{XUV}}{L_{XUV\odot}}\big)^\mu     ~~~\text{mho},
\label{eq:Ped}
\end{equation}

\noindent where $\kappa = 15.475$, $\lambda = -2.082$, $\mu = 1/2$, $d$ is orbital distance and $L_{XUV}$ is stellar X-ray and Extreme ultraviolet (EUV; together, XUV) \citep{nichols2016stellar}.  For simplicity, we assume the Hall conductance is zero for all simulations.\\ %\citep{} e.g. Fedder & Lyon 1987; Jia, Kivelson & Gombosi 2012a; Jia et al. 2012b).\\

\begin{figure}[ht]
%\plotone{qkpic04au.png}
\centering
\includegraphics[width=85mm]{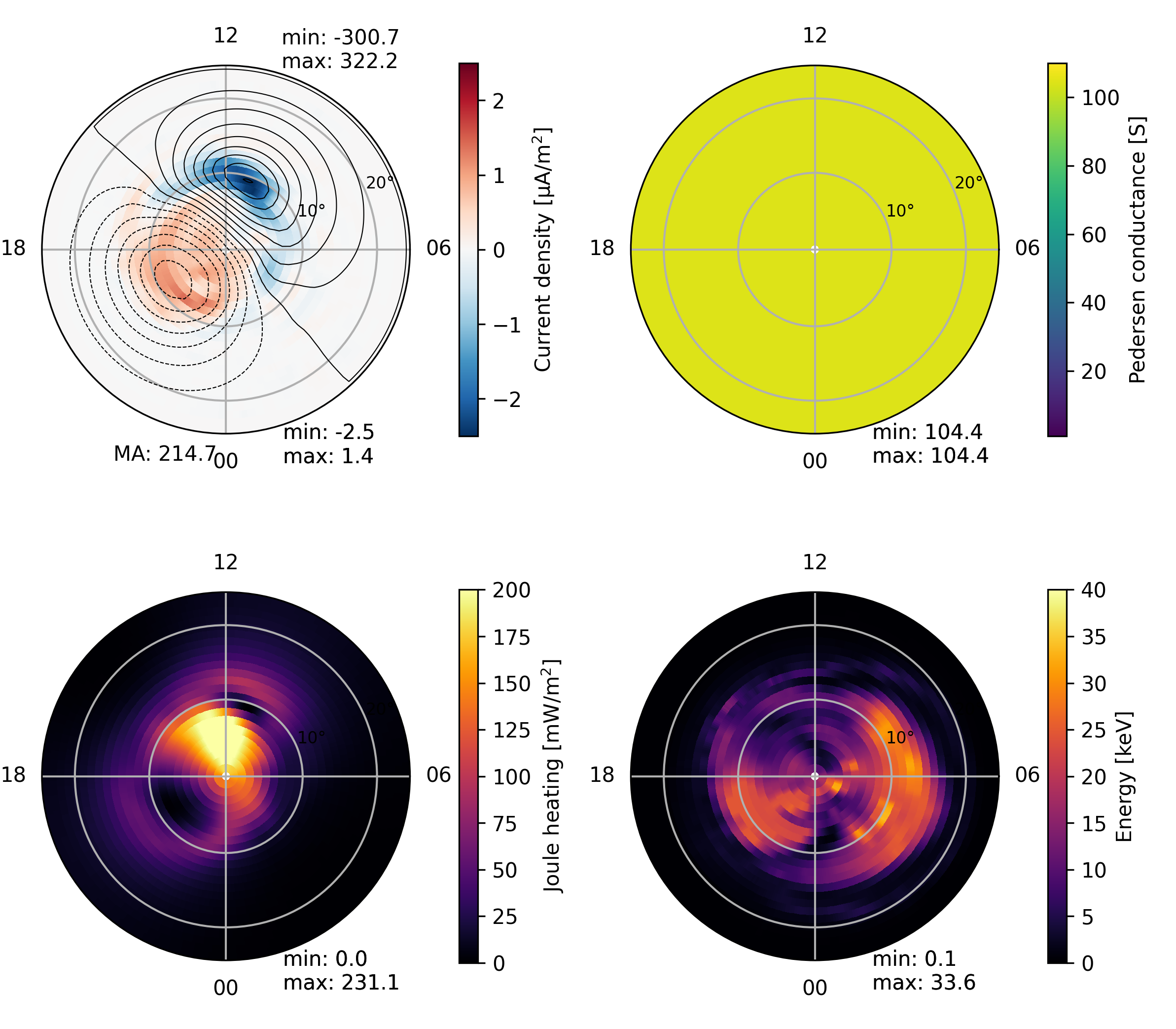}
\includegraphics[width=85mm]{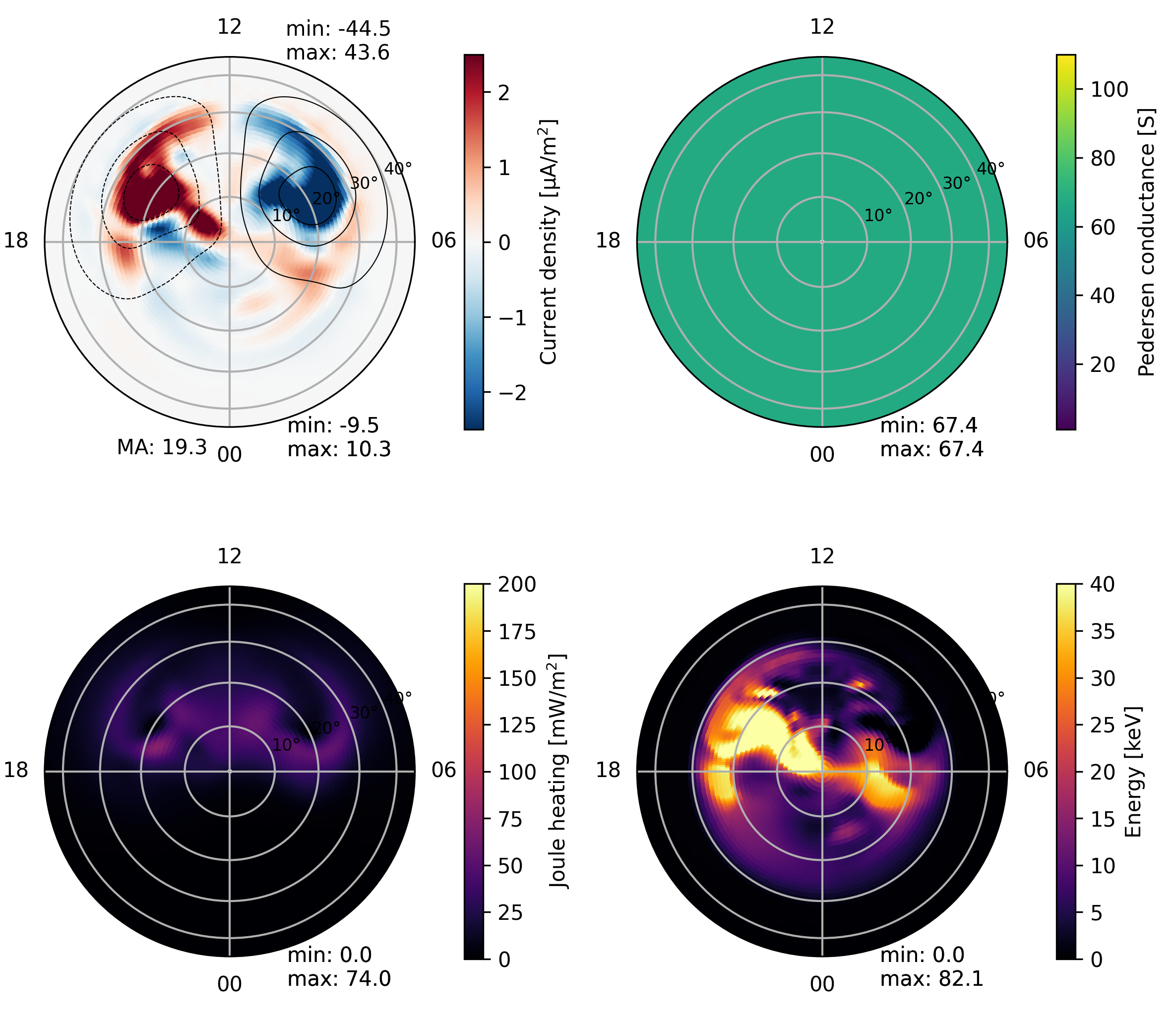}
\caption{ The simulations output for current density, Joule heating, and energy, the northern hemisphere of \textit{left}- a Jovian and \textit{right}- a rocky planet, both at 0.4 au with the IMF = -32 nT. \label{fig:04auRot_1}}
\end{figure}

\noindent Figure \ref{fig:04auRot_1} illustrates the simulation outcomes depicting FACs, Joule heating, energy distribution, and Pedersen Conductance for both an Earth-like rocky planet and a Jovian Planet. The right panel of Figure \ref{fig:04auRot_1} showcases the northern hemisphere's FACs, energy flux, and Joule heating distribution. In this simulation, the Pedersen conductance, calculated using Eq. \ref{eq:Ped}, was $\Sigma_P = 104.4$ mho, while the Hall conductance was set to zero, as depicted in Figure \ref{fig:04auRot_1}. As anticipated, all energy components within the magnetosphere-ionosphere system exhibit larger magnitudes for the Jovian planet, attributable to its significantly stronger magnetic field.
\\

\section{Tidally locked exoplanets}

\begin{figure}[ht]
%\plotone{qkpic04au.png}
\centering
\includegraphics[width=100mm]{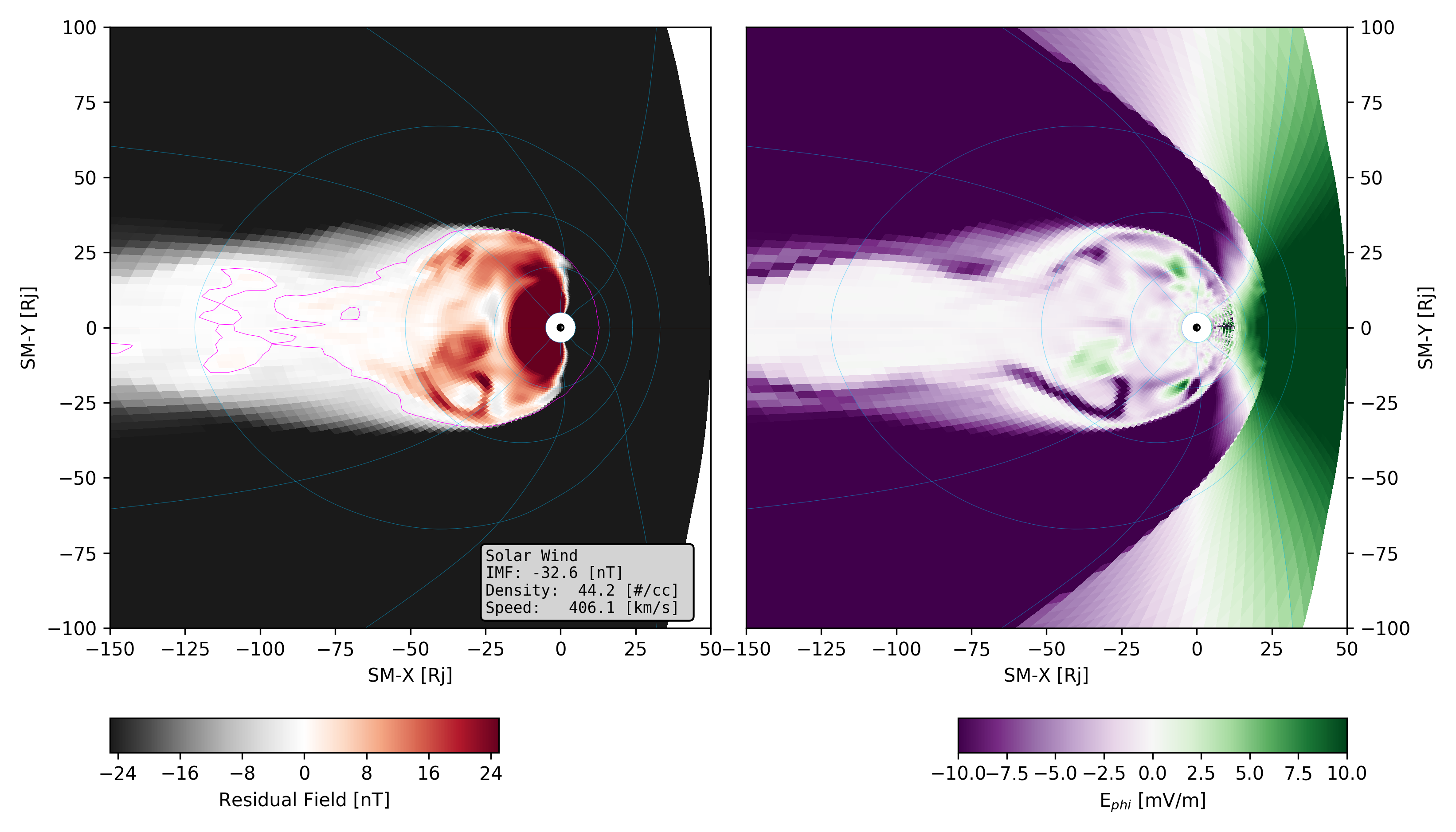}
\includegraphics[width=70mm]{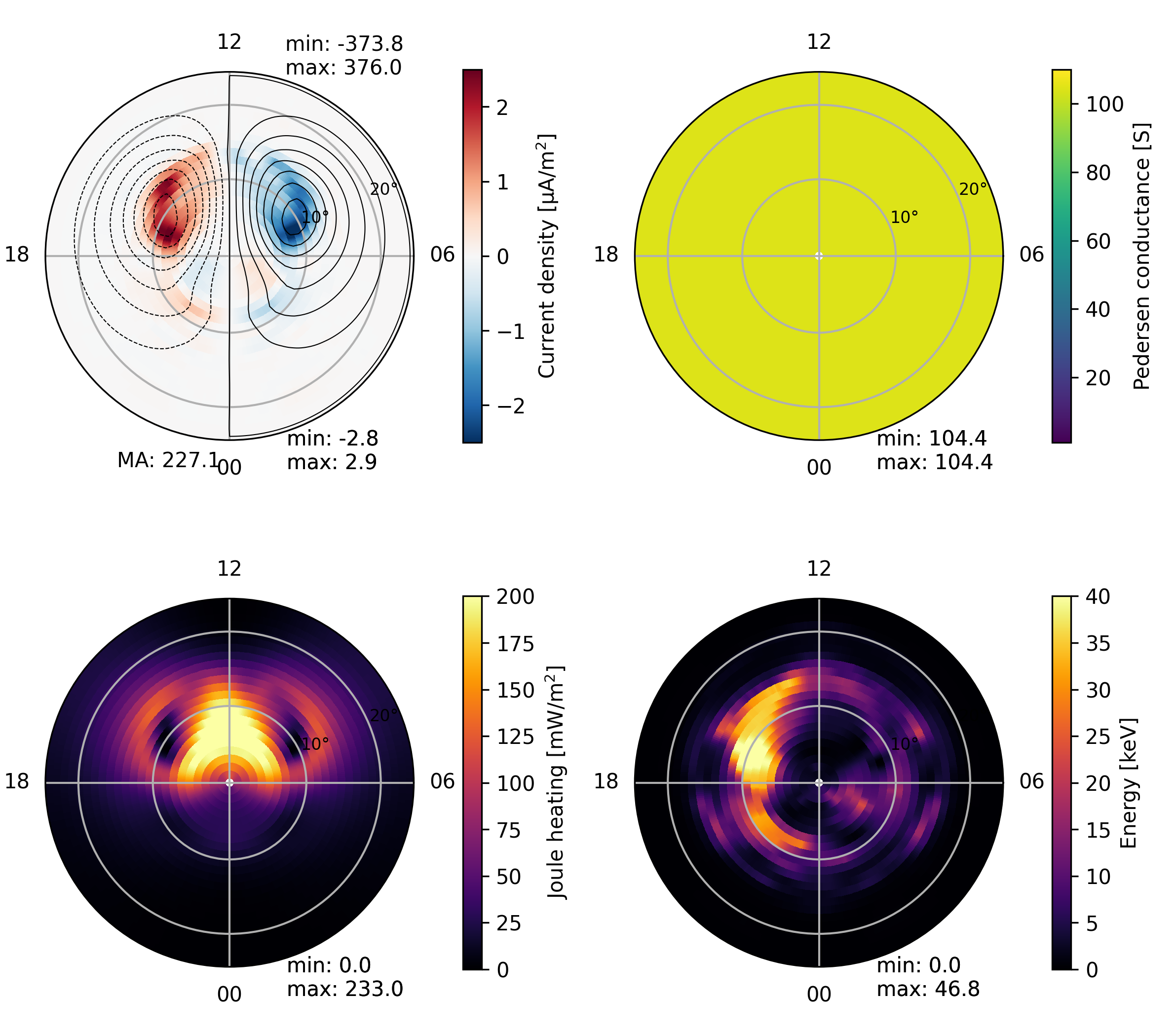}
\caption{\textit{Left}- Azimuthal component of electric field and Residual Field for a tidally-locked Jupiter-like planet at 0.4 au with the IMF = -32 nT. \textit{Right}- The simulations output for current density, Joule heating, and energy in the northern hemisphere, a tidally-locked Jupiter-like planet at 0.4 au.
\label{fig:04auNoRot_1}}
\end{figure}

\noindent A crucial factor in estimating the radio flux of exoplanets is the planetary magnetic moment, a quantity assumed in MHD simulations. Two theoretical approaches underlie its calculation \citep{weber2017expanded, farrell1999possibility}. The first relies on scaling laws, linking the planetary magnetic dipole moment to its rotational speed \citep{griessmeier2004effect}, indicating a direct influence of planetary rotations on the magnetic field. Conversely, the second hypothesis attributes the planetary magnetic moment primarily to energy flux from the planetary core \citep{reiners2010magnetic}, largely unaffected by tidal locking. Tidal locking, a consequence of strong tidal dissipation in exoplanets orbiting at smaller distances, significantly impacts the generation of auroral radiation due to asymmetrical conductance in the ionosphere \citep{zarka2001magnetically, seager2002constraining}. Consequently, to model exoplanetary radio emissions accurately, MHD simulations must consider the effects of tidal locking and exoplanet rotation, critical factors that have yet been unexplored in previous studies. Notably, a planet at 0.4 au would become tidally locked much sooner than one at 1 au or beyond. A slower rotation rate may weaken the dynamo effect, resulting in weaker magnetic fields and smaller magnetospheres. However, recent studies \citep{zuluaga2013influence} show a non-trivial relationship between rotation period and magnetic properties, suggesting that tidally locked planets could still exhibit intense magnetic fields and extended magnetospheres with larger polar cap areas.\\ 
\\
In this section, we aim to explore the influence of tidal locking on magnetospheric dynamics and its interaction with stellar winds. Specifically, we seek to elucidate the impact of rotation speed on magnetospheric energy flow, comparing two scenarios: one featuring a rapidly rotating planet akin to Jupiter and the other exhibiting synchronous rotation matching its orbital angular velocity. To address this inquiry, we conduct simulations mirroring the preceding section, focusing on a Jovian planet positioned at 0.4 au but assuming a rotational period equivalent to its orbital period. Changing the rotation period means changing the corotation potential in the magnetosphere-ionosphere system. Basically, corotation potential denotes the electrostatic potential arising from the dynamic interplay between a planet's magnetosphere and the surrounding plasma environment. This phenomenon manifests prominently in regions where the planet's rotational velocity synchronizes with the flow of the surrounding plasma, particularly in the outer reaches of the magnetosphere. In such zones of corotation, the plasma particles co-rotate with the planet's magnetic field lines, resulting in a characteristic distribution of electrostatic potential. This potential can be driven by
\begin{equation}
\Phi_{\text{corotation}} = - \Omega~B~ R \sin{\lambda}~,
\end{equation}
where $\Omega$ is the angular speed of Planet's rotation, $B$ is the equatorial magnetic field strength, $R$ is planetary radius, and 
$\lambda$ is the magnetic co-latitude.
Regions exhibiting corotation potential are characterized by relatively low potential energy, indicating a harmonized motion between the planet and the surrounding plasma. Conversely, areas where the plasma's velocity deviates from the planet's rotational motion experience heightened corotation potential.\\
\\
The simulation outcomes, detailing the characteristics of magnetospheric magnetic and electric fields, are illustrated in Figure \ref{fig:04auNoRot_1}. The right panel of Figure \ref{fig:04auNoRot_1} illustrates current density, Joule heating, energy distribution, and Pedersen conductance within the ionosphere-magnetosphere system of a tidally-locked Jovian planet at 0.4 au. A comparison between Figures \ref{fig:04auNoRot_1} and \ref{fig:04auRot_1} reveals that in the tidally-locked scenario, FACs exhibit a symmetric pattern and higher magnitude attributable to reduced heating dissipation.

\begin{figure}[ht]
%\plotone{qkpic04au.png}
\centering
\includegraphics[width=80mm]{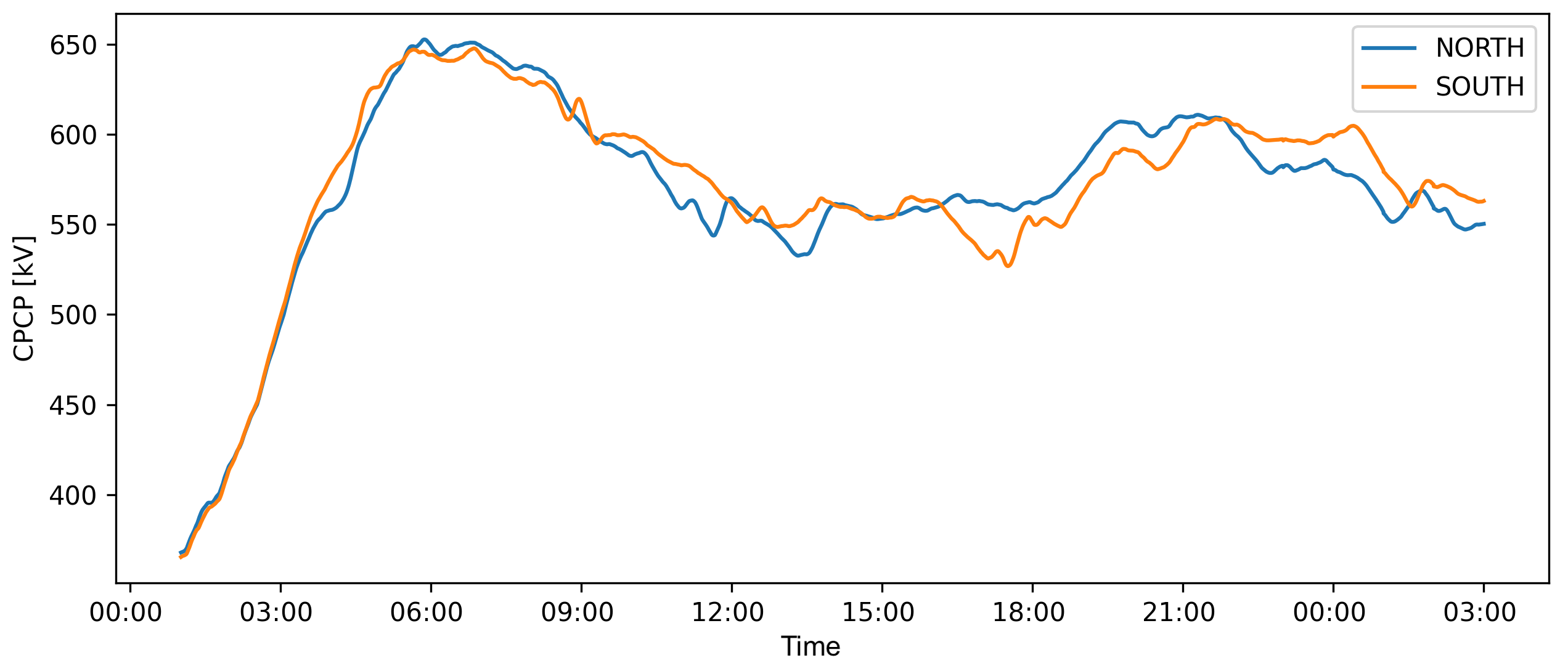}
\includegraphics[width=80mm]{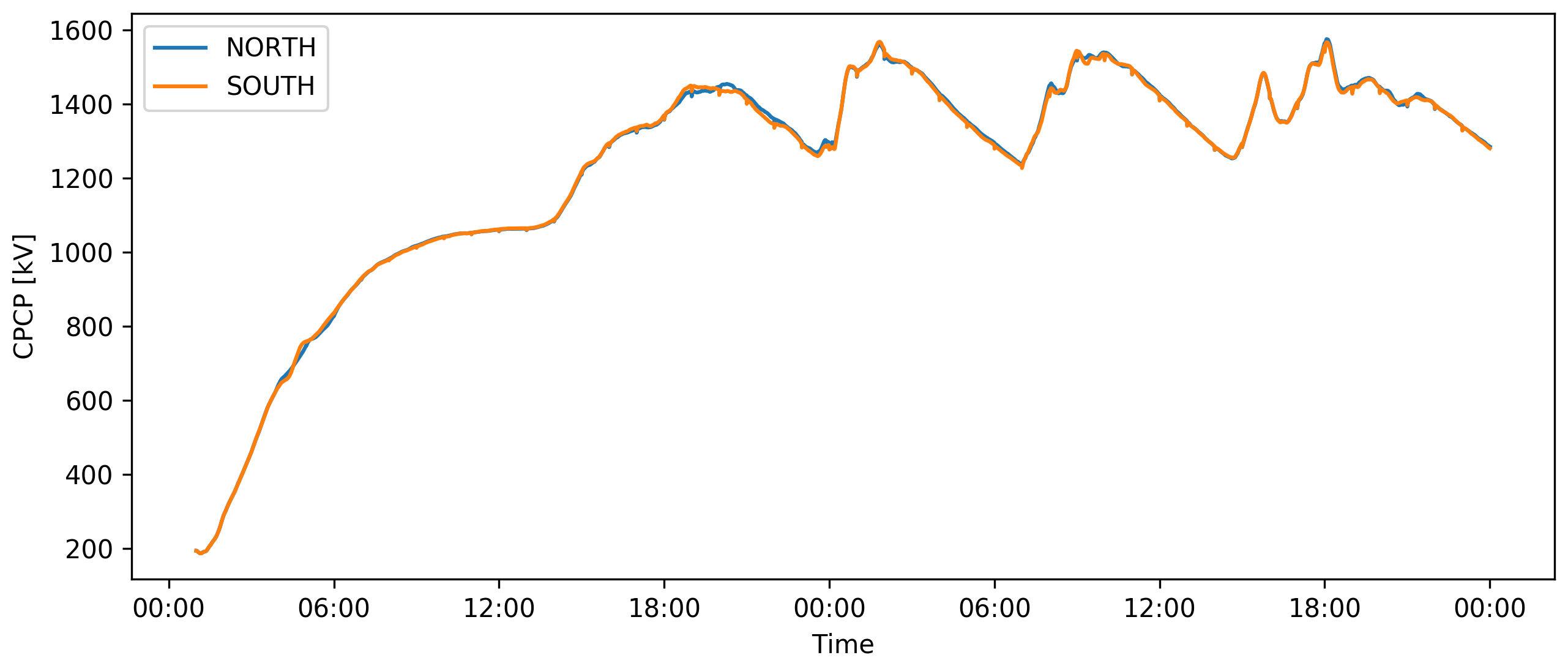}
\caption{Cross Polar Cap Potential in the northern and southern hemispheres for \textit{left}- a fast rotating planet (10 hours rotation period), and \textit{right}- a tidally locked planet. 
\label{fig:cpcp}}
\end{figure}

\begin{figure}[ht]
%\plotone{qkpic04au.png}
\centering
\includegraphics[width=85mm]{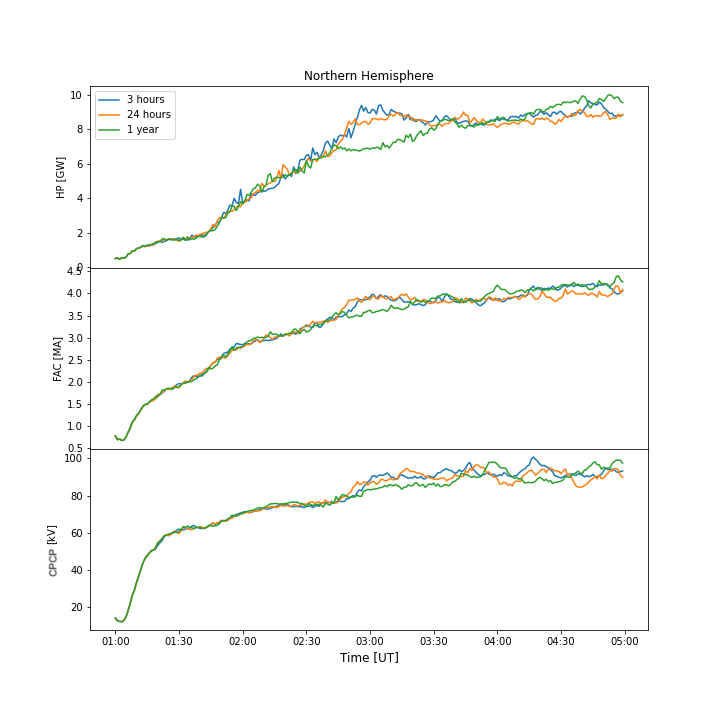}
\includegraphics[width=85mm]{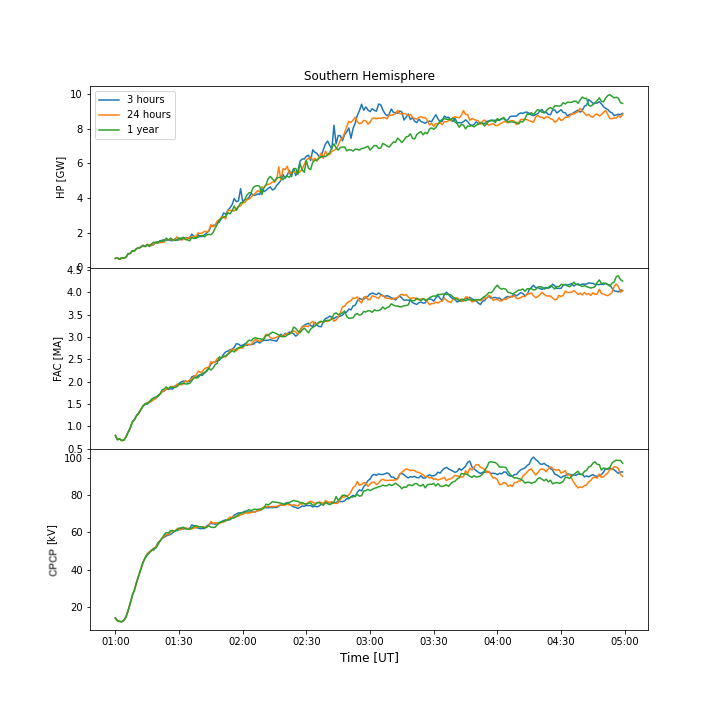}
\caption{HP, FACs, and Cross potential polar cap with different corotation periods in \textit{left}- northern and \textit{right}- southern hemisphere. %In these simulations, the orbital distance is 1 au and the typical value for solar wind is assumed. 
\label{fig:rockycpcp}}
\end{figure}

\noindent One of the principal distinctions observed in the context of tidal locking pertains to the evolution of the Cross-Polar Cap Potential (CPCP) throughout the simulation. Illustrated in the left panel of Figure \ref{fig:cpcp}, the CPCP for a rapidly rotating planet reaches its zenith approximately five hours into the simulation period. Conversely, in the scenario featuring a tidally locked planet (right panel), the CPCP exhibits continual escalation, ultimately peaking at approximately twice that of the CPCP observed in the fast-rotating planet configuration with an equivalent magnetic field. This trend can be understood by considering that tidally locked planets possess an expanded polar cap region, which in turn contributes to higher CPCP \citep{zuluaga2013influence}.\\  
\\
\noindent We also explore the influence of corotation on the magnetospheric dynamics of a rocky planet possessing an Earth-like magnetic field. Figure \ref{fig:rockycpcp} depicts CPCP, FACs, and Hemispheric Power (HP) of the northern (left) and southern hemispheres (right), considering various corotation periods. Although there are some differences, the overall pattern remains similar for all rotation speeds, unlike what we see with Jovian planets. A comprehensive analysis is needed to explain the observed elevation in CPCP in simulations of tidally-locked Jovian planets. However, this falls beyond the scope of the present study and is undergoing comprehensive investigation in a subsequent paper.

\section{Orbital distance effects}

\begin{figure}[ht]
%\plotone{qkpic04au.png}
\centering
\includegraphics[width=85mm]{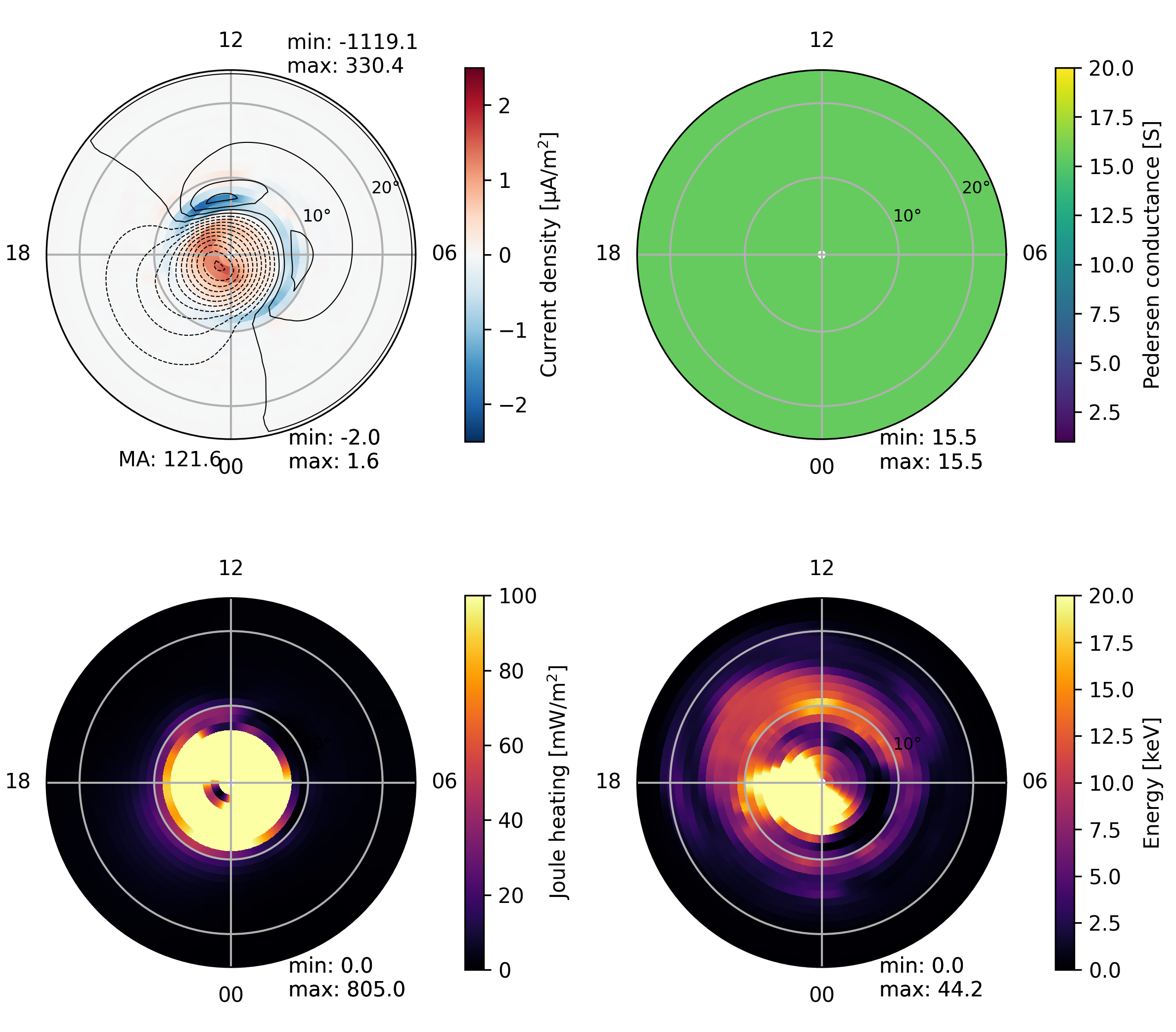}
\includegraphics[width=85mm]{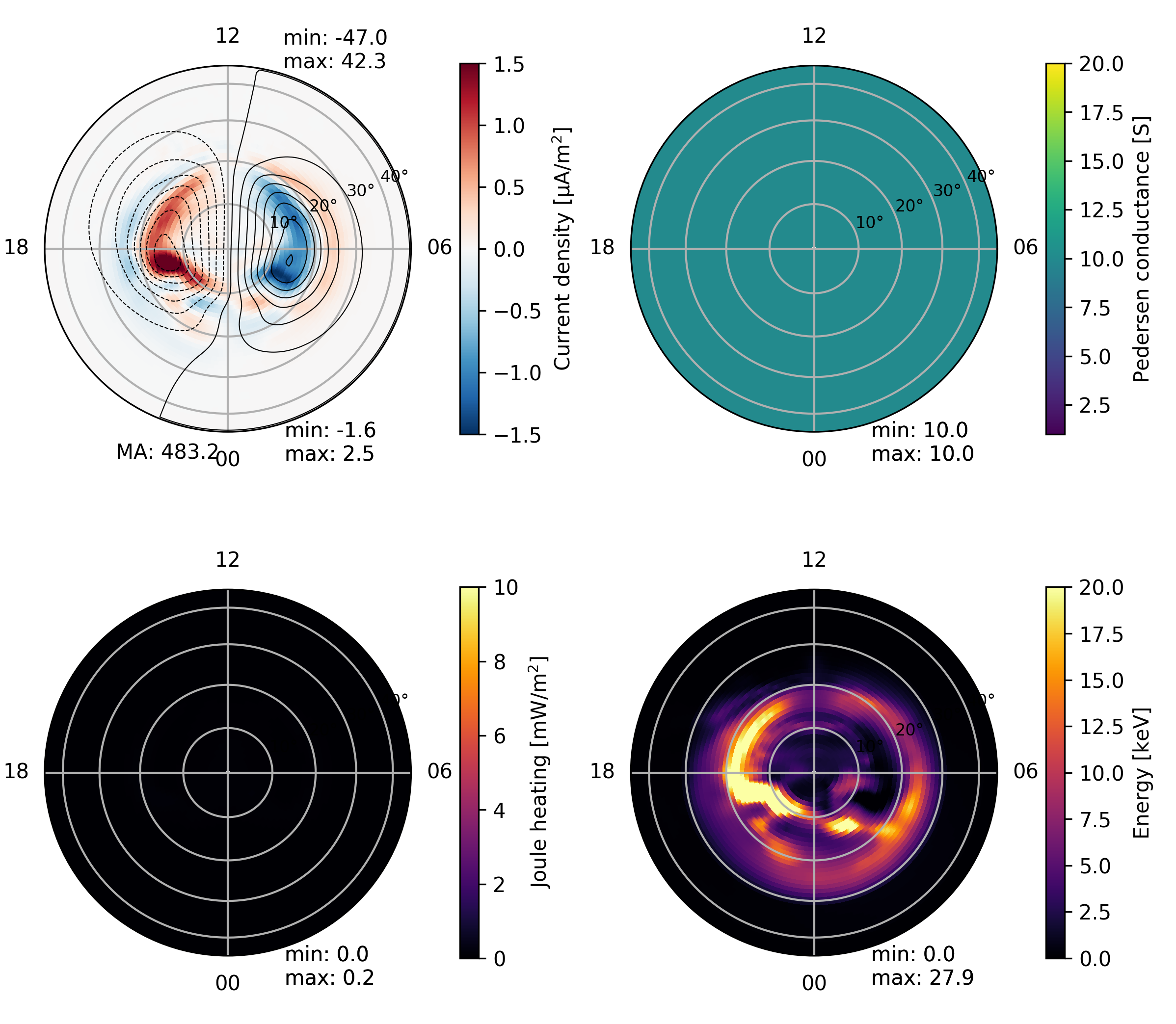}
\caption{The simulations output for current density, Joule heating, and energy, the northern hemisphere of \textit{left}- a Jovian and \textit{right}- a rocky planet at 1 au with a Sun-like star.
\label{fig:1auRot_1}}
\end{figure}

\noindent To investigate the influence of orbital distance on the magnetosphere of planets, we replicate the experimental setup with planets positioned at 1 au. Considering a Sun-like star, the stellar wind parameters at this distance are typical values observed near Earth. These include an incident plasma velocity of 400 $km/s$, a plasma temperature of 1 MK, a density of 5 $cc^{-1}$, a dynamic pressure of 1.3 nPa, and an interplanetary magnetic field (IMF) typically around -5 nT in non-storm conditions. As illustrated in the left panel of Figure \ref{fig:1auRot_1}, for a Jovian planet, the calculated Pedersen conductance from equation \ref{eq:Ped} is 104.4 mho. Conversely, the right panel illustrates the same parameters for a rocky planet at 1 au. The rotation period for the rocky planet is 24 hours, while for the Jovian planet, it stands at 10 hours. The Joule heating and FACs in the Jovian planet's magnetosphere are approximately one order of magnitude greater than those observed in the rocky planet's magnetosphere. Consequently, the Auroral Kilometric Radiation (AKR) emission from the Jovian planet is anticipated to be at least one magnitude more intense than that of an Earth-like planet.

\section{An exoplanet with an M Dwarf star}

\begin{figure}[ht!]
%\plotone{qkpic04au.png}
%\includegraphics[width=100mm]{MDwarf_CoRot_31hrs_1au_qkpic.png}
\centering
\includegraphics[width=85mm]{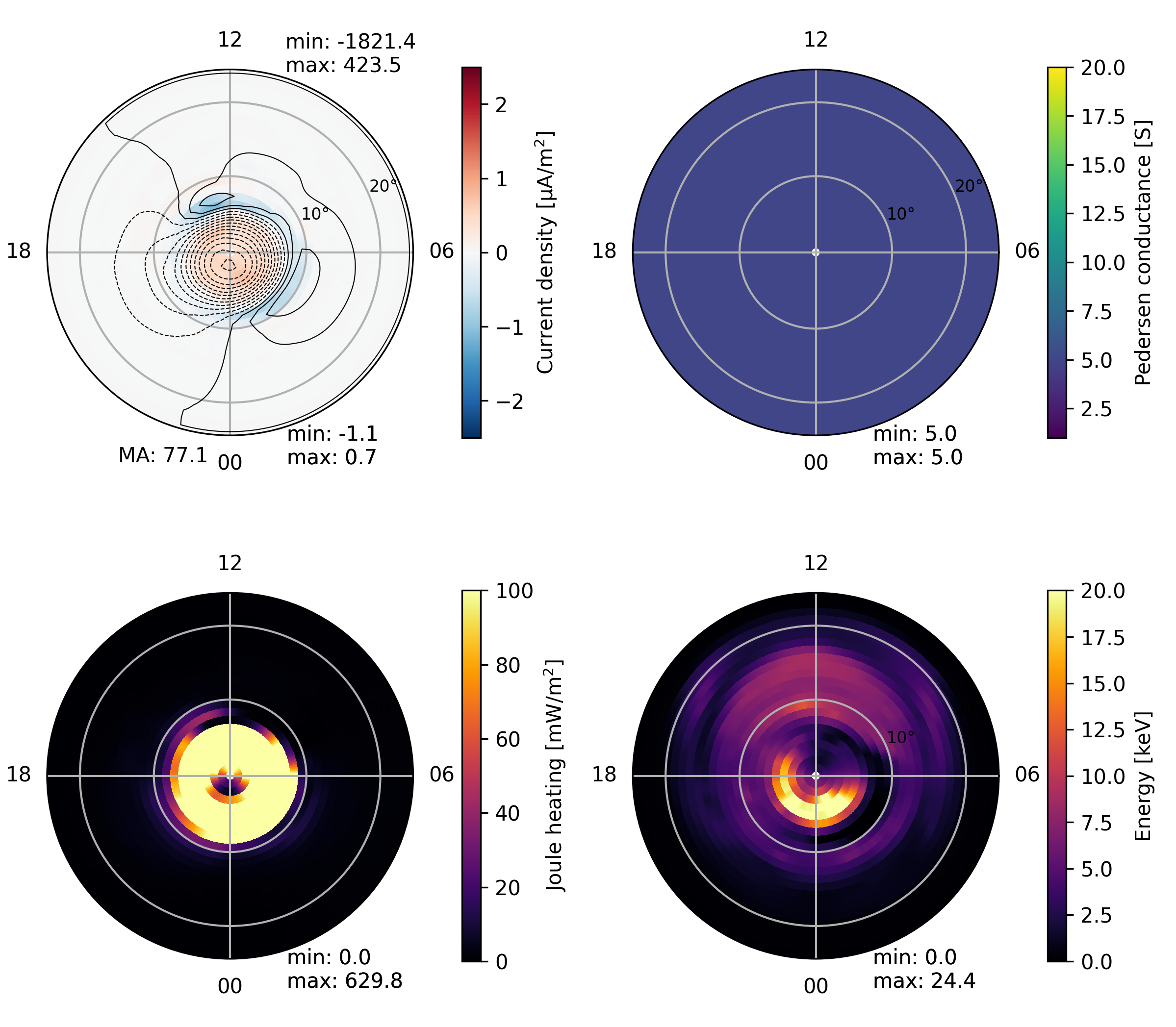}
\includegraphics[width=85mm]{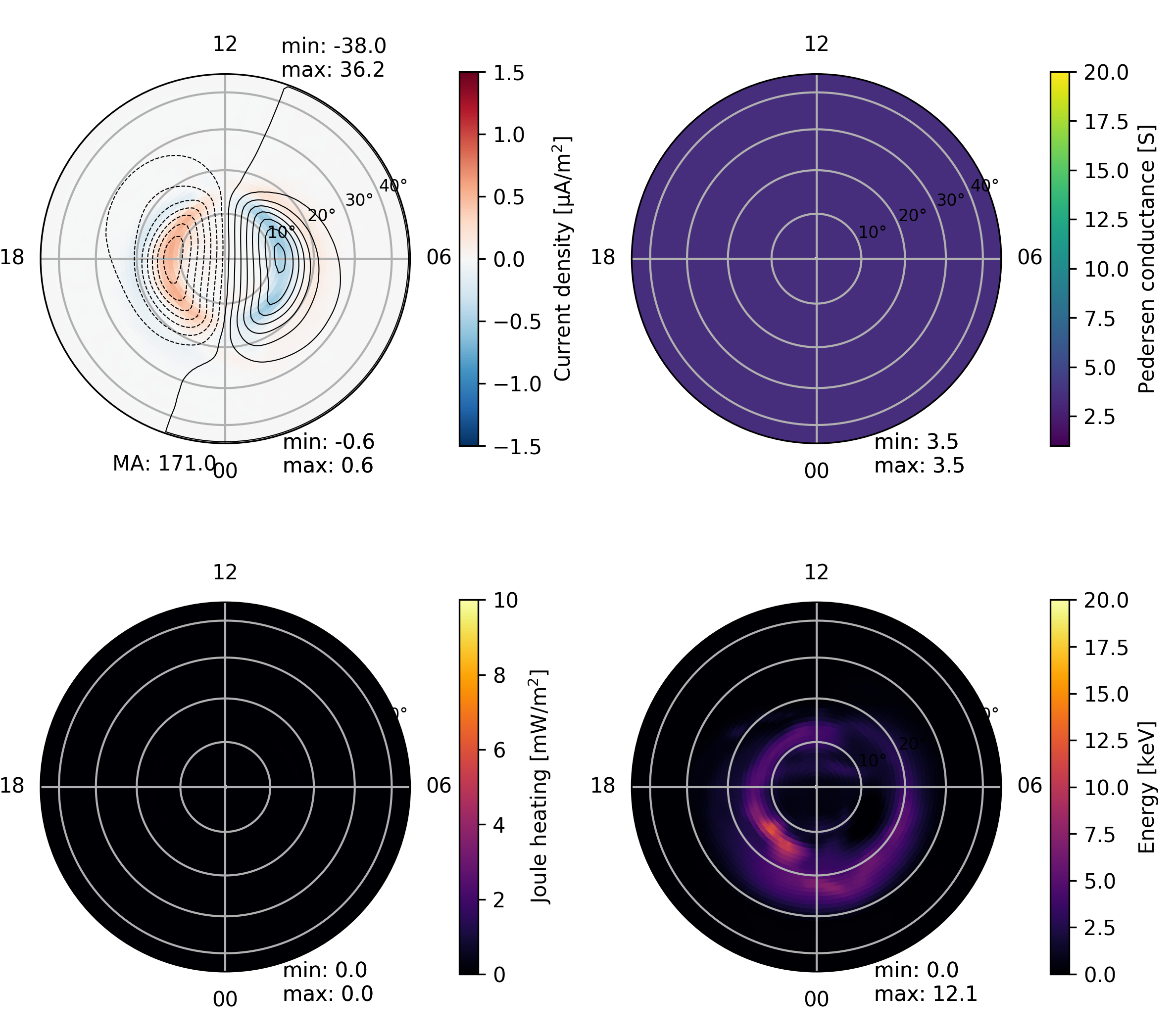}

\caption{The simulations output for current density, Joule heating, and energy, the northern hemisphere of \textit{left}- a Jovian and \textit{right}- a rocky planet at 1 au with an M Dwarf star
\label{fig:1auMdwarf}}
\end{figure}

\noindent To contrast the preceding outcomes with the planetary interaction under the influence of a cooler stellar body, we conduct a simulation using standard values for M dwarf stellar wind parameters. In the scenario involving an M dwarf star, the stellar wind parameters at 1 au encompass an incident plasma velocity of 370 $km/s$, a plasma temperature of 0.41 MK, a density of 1 $cc^{-1}$, a dynamic pressure of 0.23 nPa, and an IMF typically around -2 nT in non-storm conditions \citep{fleming2020xuv, vidotto2017exoplanets}. Given that M dwarfs emit less XUV radiation compared to Sun-like stars, the calculated Pedersen conductance from equation \ref{eq:Ped} stands at $\Sigma_P = 5.0$ mho for a Jovian planet and $\Sigma_P = 3.5$ for a rocky planet in this simulation. As shown in Figure \ref{fig:1auMdwarf}, the total current in this context is roughly half that observed in scenarios involving Sun-like stars.

\section{Conclusion}

\noindent As previously mentioned, current 3D MHD simulations like GAMERA/LFM or SWMF are constrained by assumptions based on the solar system and Earth, rendering them inadequate for modeling systems with close-in exoplanets. Consequently, we cannot position a planet with a Jupiter-sized (or larger) magnetic field closer than 0.4 au to its host star. The magnetospheric structure of close-in exoplanets markedly differs from that of solar system planets. Close-in exoplanets tend to possess shock-less magnetospheres and are typically situated within the sub-Alfv\'enic zone of their host stars. Their proximity to host stars and relatively large sizes often fill their Roche lobes, leading to outflows from Lagrange points L1 and L2. The close proximity of these exoplanets to their host stars causes intense heating, ionization, and chemical alteration of their upper atmospheres due to stellar X-ray/EUV radiation, prompting hydrodynamic expansion of ionized atmospheric material. Consequently, modeling the magnetosphere-ionosphere interactions of exoplanets necessitates substantial modifications to established MHD models.\\ 
\\
In this study, we position a planet at a distance of 0.4 au to ensure a reasonably realistic simulation scenario. We explored the impact of tidal locking on the system dynamics. The observed rise in CPCP in simulations of tidally-locked planets warrants further examination, which we plan to address in a subsequent manuscript. It is essential to acknowledge that conductance is variable in actual conditions, particularly for tidally locked planets. These findings represent an initial step toward understanding such star-planet systems.\\
\\
Moreover, we change the orbital distance from 0.4 to 1 au. When considering a Sun-like star, the conductance of a planet positioned at an orbital distance of 0.4 au surpasses that at 1 au by approximately an order of magnitude. This differential conductance results in diminished CPCP \citep{lopez2010role}, elevated total power, and heightened FACs for the planet located at 0.4 au relative to 1 au.\\
\\
Furthermore, the host stars of exoplanets may exhibit a range of spectral types, extending from F to M.  Hence, we consider two scenarios, one with a sun-like star and another with an M dwarf stellar wind, to contrast simulation outcomes. Notably, the total power within the planet's ionosphere-magnetosphere system is amplified when under the influence of a sun-like star. This amplification stems from the planet's conductance being approximately threefold greater than the M dwarf scenario. Consequently, the CPCP is correspondingly lower in the case of the sun-like star.

\section*{acknowledgments}
\noindent We acknowledge the use of the Cheyenne supercomputer at the National Center for Atmospheric Research (NCAR) \citep{computational2017cheyenne}. This research is supported by NSF grant number 2138122.
%\end{acknowledgments}

\bibliographystyle{aasjournal} %  Many Frontiers journals use the Harvard referencing system (Author-date), to find the style and resources for the journal you are submitting to: https://zendesk.frontiersin.org/hc/en-us/articles/360017860337-Frontiers-Reference-Styles-by-Journal. For Humanities and Social Sciences articles please include page numbers in the in-text citations
\bibliography{main}

\begin{thebibliography}{}
\expandafter\ifx\csname natexlab\endcsname\relax\def\natexlab#1{#1}\fi
\providecommand{\url}[1]{\href{#1}{#1}}
\providecommand{\dodoi}[1]{doi:~\href{http://doi.org/#1}{\nolinkurl{#1}}}
\providecommand{\doeprint}[1]{\href{http://ascl.net/#1}{\nolinkurl{http://ascl.net/#1}}}
\providecommand{\doarXiv}[1]{\href{https://arxiv.org/abs/#1}{\nolinkurl{https://arxiv.org/abs/#1}}}

\bibitem[{Asmar {et~al.}(2021)Asmar, Preston, Vergados, Atkinson, Andert, Ando, Ao, Armstrong, Ashby, Barriot, {et~al.}}]{asmar2021solar}
Asmar, S., Preston, R., Vergados, P., {et~al.} 2021, Bulletin of the AAS, 53

\bibitem[{Bagheri \& Lopez(2022)}]{bagheri2022solar}
Bagheri, F., \& Lopez, R.~E. 2022, Frontiers in Astronomy and Space Sciences, 9, 960535

\bibitem[{Bagheri {et~al.}(2019)Bagheri, Sajadian, \& Rahvar}]{bagheri2019detection}
Bagheri, F., Sajadian, S., \& Rahvar, S. 2019, Monthly Notices of the Royal Astronomical Society, 490, 1581

\bibitem[{Cauley {et~al.}(2019)Cauley, Shkolnik, Llama, \& Lanza}]{cauley2019magnetic}
Cauley, P.~W., Shkolnik, E.~L., Llama, J., \& Lanza, A.~F. 2019, Nature Astronomy, 3, 1128

\bibitem[{Cohen {et~al.}(2022)Cohen, Alvarado-G{\'o}mez, Drake, Harbach, Garraffo, \& Fraschetti}]{cohen2022space}
Cohen, O., Alvarado-G{\'o}mez, J.~D., Drake, J.~J., {et~al.} 2022, The Astrophysical Journal, 934, 189

\bibitem[{Cohen {et~al.}(2014)Cohen, Drake, Glocer, Garraffo, Poppenhaeger, Bell, Ridley, \& Gombosi}]{cohen2014magnetospheric}
Cohen, O., Drake, J., Glocer, A., {et~al.} 2014, The Astrophysical Journal, 790, 57

\bibitem[{Cohen {et~al.}(2011)Cohen, Kashyap, Drake, Sokolov, Garraffo, \& Gombosi}]{cohen2011dynamics}
Cohen, O., Kashyap, V., Drake, J., {et~al.} 2011, The Astrophysical Journal, 733, 67

\bibitem[{Computational \& Laboratory(2017)}]{computational2017cheyenne}
Computational, \& Laboratory, I.~S. 2017, Cheyenne: HPE/SGI ICE XA system (climate simulation laboratory),  National Center for Atmospheric Research Boulder, CO

\bibitem[{Desch(1982)}]{desch1982evidence}
Desch, M.~D. 1982, Journal of Geophysical Research: Space Physics, 87, 4549

\bibitem[{Diego {et~al.}(2020)Diego, Piersanti, Laurenza, \& Villante}]{diego2020properties}
Diego, P., Piersanti, M., Laurenza, M., \& Villante, U. 2020, Journal of Geophysical Research: Space Physics, 125, e2020JA028281

\bibitem[{Farrell {et~al.}(1999)Farrell, Desch, \& Zarka}]{farrell1999possibility}
Farrell, W., Desch, M., \& Zarka, P. 1999, Journal of Geophysical Research: Planets, 104, 14025

\bibitem[{Fleming {et~al.}(2020)Fleming, Barnes, Luger, \& VanderPlas}]{fleming2020xuv}
Fleming, D.~P., Barnes, R., Luger, R., \& VanderPlas, J.~T. 2020, The Astrophysical Journal, 891, 155

\bibitem[{Freeman \& Lopez(1985)}]{freeman1985cold}
Freeman, J.~W., \& Lopez, R.~E. 1985, Journal of Geophysical Research: Space Physics, 90, 9885

\bibitem[{Gallagher \& D'angelo(1981)}]{gallagher1981correlations}
Gallagher, D.~L., \& D'angelo, N. 1981, Geophysical Research Letters, 8, 1087

\bibitem[{Garcia-Sage {et~al.}(2023)Garcia-Sage, Farrish, Airapetian, Alexander, Cohen, Domagal-Goldman, Dong, Gronoff, Halford, Lazio, {et~al.}}]{garcia2023star}
Garcia-Sage, K., Farrish, A., Airapetian, V., {et~al.} 2023

\bibitem[{Gaudi(2012)}]{gaudi2012microlensing}
Gaudi, B.~S. 2012, Annual Review of Astronomy and Astrophysics, 50, 411

\bibitem[{Genova {et~al.}(1989)Genova, Zarka, \& Lecacheux}]{genova1989jupiter}
Genova, F., Zarka, P., \& Lecacheux, A. 1989, NASA Special Publication, 494

\bibitem[{Grie$\beta$meier {et~al.}(2005)Grie$\beta$meier, Stadelmann, Motschmann, Belisheva, Lammer, \& Biernat}]{griebetameier2005cosmic}
Grie$\beta$meier, J.-M., Stadelmann, A., Motschmann, U., {et~al.} 2005, Astrobiology, 5, 587

\bibitem[{Grie{\ss}meier {et~al.}(2007{\natexlab{a}})Grie{\ss}meier, Preusse, Khodachenko, Motschmann, Mann, \& Rucker}]{griessmeier2007exoplanetary}
Grie{\ss}meier, J.-M., Preusse, S., Khodachenko, M., {et~al.} 2007{\natexlab{a}}, Planetary and Space Science, 55, 618

\bibitem[{Grie{\ss}meier {et~al.}(2007{\natexlab{b}})Grie{\ss}meier, Zarka, \& Spreeuw}]{griessmeier2007predicting}
Grie{\ss}meier, J.-M., Zarka, P., \& Spreeuw, H. 2007{\natexlab{b}}, Astronomy \& Astrophysics, 475, 359

\bibitem[{Grie{\ss}meier {et~al.}(2004)Grie{\ss}meier, Stadelmann, Penz, Lammer, Selsis, Ribas, Guinan, Motschmann, Biernat, \& Weiss}]{griessmeier2004effect}
Grie{\ss}meier, J.-M., Stadelmann, A., Penz, T., {et~al.} 2004, Astronomy \& Astrophysics, 425, 753

\bibitem[{Gronoff {et~al.}(2020)Gronoff, Arras, Baraka, Bell, Cessateur, Cohen, Curry, Drake, Elrod, Erwin, {et~al.}}]{gronoff2020atmospheric}
Gronoff, G., Arras, P., Baraka, S., {et~al.} 2020, Journal of Geophysical Research: Space Physics, 125, e2019JA027639

\bibitem[{Guyon {et~al.}(2005)Guyon, Pluzhnik, Galicher, Martinache, Ridgway, \& Woodruff}]{guyon2005exoplanet}
Guyon, O., Pluzhnik, E.~A., Galicher, R., {et~al.} 2005, The Astrophysical Journal, 622, 744

\bibitem[{Harbach {et~al.}(2021)Harbach, Moschou, Garraffo, Drake, Alvarado-G{\'o}mez, Cohen, \& Fraschetti}]{harbach2021stellar}
Harbach, L.~M., Moschou, S.~P., Garraffo, C., {et~al.} 2021, The Astrophysical Journal, 913, 130

\bibitem[{Ip {et~al.}(2004)Ip, Kopp, \& Hu}]{ip2004star}
Ip, W.-H., Kopp, A., \& Hu, J.-H. 2004, The Astrophysical Journal, 602, L53

\bibitem[{Jackson {et~al.}(2008)Jackson, Greenberg, \& Barnes}]{jackson2008tidal}
Jackson, B., Greenberg, R., \& Barnes, R. 2008, The Astrophysical Journal, 678, 1396

\bibitem[{Jia \& Kivelson(2021)}]{jia2021magnetosphere}
Jia, X., \& Kivelson, M.~G. 2021, Magnetospheres in the solar system, 557

\bibitem[{Johnstone {et~al.}(2019)Johnstone, Khodachenko, L{\"u}ftinger, Kislyakova, Lammer, \& G{\"u}del}]{johnstone2019extreme}
Johnstone, C., Khodachenko, M., L{\"u}ftinger, T., {et~al.} 2019, Astronomy \& Astrophysics, 624, L10

\bibitem[{Johnstone {et~al.}(2018)Johnstone, G{\"u}del, Lammer, \& Kislyakova}]{johnstone2018upper}
Johnstone, C.~P., G{\"u}del, M., Lammer, H., \& Kislyakova, K.~G. 2018, Astronomy \& Astrophysics, 617, A107

\bibitem[{Knutson {et~al.}(2014)Knutson, Benneke, Deming, \& Homeier}]{knutson2014featureless}
Knutson, H.~A., Benneke, B., Deming, D., \& Homeier, D. 2014, Nature, 505, 66

\bibitem[{Koskinen {et~al.}(2013)Koskinen, Harris, Yelle, \& Lavvas}]{koskinen2013escape}
Koskinen, T., Harris, M., Yelle, R., \& Lavvas, P. 2013, Icarus, 226, 1678

\bibitem[{Laneuville {et~al.}(2020)Laneuville, Dong, O’Rourke, \& Schneider}]{laneuville2020magnetic}
Laneuville, M., Dong, C., O’Rourke, J.~G., \& Schneider, A.~C. 2020, in Planetary Diversity: Rocky planet processes and their observational signatures (IOP Publishing)

\bibitem[{Lopez {et~al.}(2010)Lopez, Bruntz, Mitchell, Wiltberger, Lyon, \& Merkin}]{lopez2010role}
Lopez, R., Bruntz, R., Mitchell, E., {et~al.} 2010, Journal of Geophysical Research: Space Physics, 115

\bibitem[{Lopez(2016)}]{lopez2016integrated}
Lopez, R.~E. 2016, Journal of Geophysical Research: Space Physics, 121, 4435

\bibitem[{Lundin {et~al.}(2007)Lundin, Lammer, \& Ribas}]{lundin2007planetary}
Lundin, R., Lammer, H., \& Ribas, I. 2007, Space Science Reviews, 129, 245

\bibitem[{Mandel \& Agol(2002)}]{mandel2002analytic}
Mandel, K., \& Agol, E. 2002, The Astrophysical Journal, 580, L171

\bibitem[{Mayor \& Queloz(1995)}]{mayor1995jupiter}
Mayor, M., \& Queloz, D. 1995, nature, 378, 355

\bibitem[{National Academies~of Sciences {et~al.}(2022)National Academies~of Sciences, Medicine, {et~al.}}]{national2022origins}
National Academies~of Sciences, E., Medicine, {et~al.} 2022

\bibitem[{Nichols \& Milan(2016)}]{nichols2016stellar}
Nichols, J., \& Milan, S. 2016, Monthly Notices of the Royal Astronomical Society, 461, 2353

\bibitem[{Ramstad \& Barabash(2021)}]{ramstad2021intrinsic}
Ramstad, R., \& Barabash, S. 2021, Space Science Reviews, 217, 1

\bibitem[{Reiners \& Christensen(2010)}]{reiners2010magnetic}
Reiners, A., \& Christensen, U.~R. 2010, Astronomy \& Astrophysics, 522, A13

\bibitem[{Saur {et~al.}(2013)Saur, Grambusch, Duling, Neubauer, \& Simon}]{saur2013magnetic}
Saur, J., Grambusch, T., Duling, S., Neubauer, F., \& Simon, S. 2013, Astronomy \& Astrophysics, 552, A119

\bibitem[{Seager \& Hui(2002)}]{seager2002constraining}
Seager, S., \& Hui, L. 2002, The Astrophysical Journal, 574, 1004

\bibitem[{Seager \& Mallen-Ornelas(2003)}]{seager2003unique}
Seager, S., \& Mallen-Ornelas, G. 2003, The Astrophysical Journal, 585, 1038

\bibitem[{Shaikhislamov {et~al.}(2020)Shaikhislamov, Khodachenko, Lammer, Berezutsky, Miroshnichenko, \& Rumenskikh}]{shaikhislamov2020three}
Shaikhislamov, I., Khodachenko, M., Lammer, H., {et~al.} 2020, Monthly Notices of the Royal Astronomical Society, 491, 3435

\bibitem[{Traub \& Oppenheimer(2010)}]{traub2010direct}
Traub, W.~A., \& Oppenheimer, B.~R. 2010, Direct imaging of exoplanets (University of Arizona Press, Tucson)

\bibitem[{Treumann(2006)}]{treumann2006electron}
Treumann, R.~A. 2006, The Astronomy and Astrophysics Review, 13, 229

\bibitem[{Turnpenney {et~al.}(2020)Turnpenney, Nichols, Wynn, \& Jia}]{turnpenney2020magnetohydrodynamic}
Turnpenney, S., Nichols, J.~D., Wynn, G.~A., \& Jia, X. 2020, Monthly Notices of the Royal Astronomical Society, 494, 5044

\bibitem[{Vidotto \& Bourrier(2017)}]{vidotto2017exoplanets}
Vidotto, A., \& Bourrier, V. 2017, Monthly Notices of the Royal Astronomical Society, 470, 4026

\bibitem[{Vidotto {et~al.}(2011)Vidotto, Llama, Jardine, Helling, \& Wood}]{vidotto2011shock}
Vidotto, A., Llama, J., Jardine, M., Helling, C., \& Wood, K. 2011, Astronomische Nachrichten, 332, 1055

\bibitem[{Weber {et~al.}(2017)Weber, Lammer, Shaikhislamov, Chadney, Khodachenko, Grie{\ss}meier, Rucker, Vocks, Macher, Odert, {et~al.}}]{weber2017expanded}
Weber, C., Lammer, H., Shaikhislamov, I.~F., {et~al.} 2017, Monthly Notices of the Royal Astronomical Society, 469, 3505

\bibitem[{Wood {et~al.}(2005)Wood, M{\"u}ller, Zank, Linsky, \& Redfield}]{wood2005new}
Wood, B.~E., M{\"u}ller, H.-R., Zank, G.~P., Linsky, J.~L., \& Redfield, S. 2005, The Astrophysical Journal, 628, L143

\bibitem[{Wright {et~al.}(2012)Wright, Marcy, Howard, Johnson, Morton, \& Fischer}]{wright2012frequency}
Wright, J., Marcy, G., Howard, A., {et~al.} 2012, The Astrophysical Journal, 753, 160

\bibitem[{Xi {et~al.}(2015)Xi, Lotko, Zhang, Brambles, Lyon, Merkin, \& Wiltberger}]{xi2015poynting}
Xi, S., Lotko, W., Zhang, B., {et~al.} 2015, Journal of Geophysical Research: Space Physics, 120, 384

\bibitem[{Zarka(2004)}]{zarka2004fast}
Zarka, P. 2004, Planetary and Space Science, 52, 1455

\bibitem[{{Zarka}(2007)}]{zarka2007plasma}
{Zarka}, P. 2007, \planss, 55, 598, \dodoi{10.1016/j.pss.2006.05.045}

\bibitem[{Zarka {et~al.}(2018)Zarka, Marques, Louis, Ryabov, Lamy, Echer, \& Cecconi}]{zarka2018jupiter}
Zarka, P., Marques, M., Louis, C., {et~al.} 2018, Astronomy \& Astrophysics, 618, A84

\bibitem[{Zarka {et~al.}(2001)Zarka, Treumann, Ryabov, \& Ryabov}]{zarka2001magnetically}
Zarka, P., Treumann, R.~A., Ryabov, B.~P., \& Ryabov, V.~B. 2001, in Physics of Space: Growth Points and Problems (Springer), 293--300

\bibitem[{Zhang {et~al.}(2019)Zhang, Sorathia, Lyon, Merkin, Garretson, \& Wiltberger}]{zhang2019gamera}
Zhang, B., Sorathia, K.~A., Lyon, J.~G., {et~al.} 2019, The Astrophysical Journal Supplement Series, 244, 20

\bibitem[{Zhilkin \& Bisikalo(2019)}]{zhilkin2019possible}
Zhilkin, A., \& Bisikalo, D. 2019, Astronomy Reports, 63, 550

\bibitem[{Zuluaga {et~al.}(2013)Zuluaga, Bustamante, Cuartas, \& Hoyos}]{zuluaga2013influence}
Zuluaga, J.~I., Bustamante, S., Cuartas, P.~A., \& Hoyos, J.~H. 2013, The Astrophysical Journal, 770, 23

\end{thebibliography}

\end{document}